\documentclass[conference]{IEEEtran}

\makeatletter
\def\ps@headings{%
\def\@oddhead{\mbox{}\scriptsize\rightmark \hfil \thepage}%
\def\@evenhead{\scriptsize\thepage \hfil \leftmark\mbox{}}%
\def\@oddfoot{}%
\def\@evenfoot{}}
\makeatother
\usepackage{amsfonts}
\usepackage{epsfig,epic}
\usepackage{url,cite}
\usepackage{amsmath,amssymb,verbatim,graphicx}
\usepackage{subfigure}
\usepackage{color}
\usepackage{multicol}

\newtheorem{prop}{Proposition}[section]

\newtheorem{cor}{Corollary}[section]

\newtheorem{lm}{Lemma}[section]

\newtheorem{thm}{Theorem}[section]

\newcommand{\bthm}{\begin{thm}}
\newcommand{\ethm}{\end{thm}}

\newcommand{\bcor}{\begin{cor}}
\newcommand{\ecor}{\end{cor}}
\newcommand{\bprop}{\begin{prop}}
\newcommand{\eprop}{\end{prop}}
\newcommand{\blm}{\begin{lm}}
\newcommand{\elm}{\end{lm}}
\newcommand{\beq}{\begin{equation}}
\newcommand{\eeq}{\end{equation}}
\newcommand{\ber}{\begin{eqnarray}}
\newcommand{\eer}{\end{eqnarray}}

\newenvironment{proof1}{\begin{trivlist}\item[]{\bf Proof:\hspace{2mm}}}{\hfill$\blackbox$\end{trivlist}}

\newcommand{\blackbox}{\vrule height7pt width5pt depth1pt}

\newcommand{\bit}{\begin{itemize}}
\newcommand{\eit}{\end{itemize}}
\newcommand{\ben}{\begin{enumerate}}
\newcommand{\een}{\end{enumerate}}
\newcommand{\bdesc}{\begin{description}}
\newcommand{\edesc}{\end{description}}
\newcommand{\beqarrn}{\begin{eqnarray*}}
\newcommand{\eeqarrn}{\end{eqnarray*}}
\newenvironment{proofof}[1]{\begin{trivlist}\item[]{\bf Proof of #1:\hspace{2mm}
}}{\hfill\blackbox\end{trivlist}}
\newcommand{\bproofof}{\begin{proofof}}
\newcommand{\eproofof}{\end{proofof}}
\newenvironment{rem}{\begin{trivlist}\item[]{\bf
Remark:}\hspace{4mm}}{\end{trivlist}}
\newcommand{\brem}{\begin{rem}}
\newcommand{\erem}{\end{rem}}
\newenvironment{rems}{\begin{trivlist}\item[]{\bf
Remarks}\begin{itemize}}{\end{itemize}\end{trivlist}}
\newcommand{\brems}{\begin{rems}}
\newcommand{\erems}{\end{rems}}
\newtheorem{fact}{Fact}
\newcommand{\bfact}{\begin{fact}}
\newcommand{\efact}{\end{fact}}
\newtheorem{examp}{Example}
\newcommand{\bexamp}{\begin{examp}\rm}
\newcommand{\eexamp}{\end{examp}}
\newtheorem{defn}{Definition}[section]
\newcommand{\bdefn}{\begin{defn}\rm}
\newcommand{\edefn}{\end{defn}}

\newtheorem{prob}{Problem}
\newcommand{\bprob}{\begin{prob}}
\newcommand{\eprob}{\end{prob}}

\newcommand{\bvtm}{\begin{verbatim}}
\newcommand{\bfig}{\begin{figure}}
\newcommand{\efig}{\end{figure}}
\newcommand{\bcen}{\begin{center}}
\newcommand{\ecen}{\end{center}}

\long\def\comment#1{}

\def \n2{{N_0 \over 2}}

\def \h5{\hspace{0.5in}}

\begin{document}

\IEEEoverridecommandlockouts

\title {{Spectrum Shaping via Network Coding in Cognitive Radio Networks}}

\author {\IEEEauthorblockN{Shanshan Wang\IEEEauthorrefmark{1},
Yalin E. Sagduyu\IEEEauthorrefmark{2}, Junshan Zhang\IEEEauthorrefmark{1} and Jason H. Li\IEEEauthorrefmark{2}}
\IEEEauthorblockA
{\IEEEauthorrefmark{1}School of Electrical, Computer and Energy Engineering,
Arizona State University, Tempe, AZ 85287, USA\\
}
\IEEEauthorblockA
{\IEEEauthorrefmark{2}Intelligent Automation, Inc., Rockville, MD 20855, USA \\
}
\thanks {This material is based upon work supported by the Air Force Office of Scientific Research under Contracts FA9550-09-C-0155 and FA9550-10-C-0026. Any opinions, findings and conclusions or recommendations expressed in this material are those of the authors and do not necessarily reflect the views of the Air Force Office of Scientific Research.}

}
\maketitle

\begin{abstract}
We consider a cognitive radio network where primary users (PUs)
employ network coding for data transmissions. We view network coding
as a spectrum shaper, in the sense that it increases spectrum
availability to secondary users (SUs) and offers more structure of
spectrum holes that improves the predictability of the primary
spectrum. With this spectrum shaping effect of network coding, each
SU can carry out adaptive channel sensing by dynamically updating
the list of the PU channels predicted to be idle while giving
priority to these channels when sensing. This dynamic spectrum
access approach with network coding improves how SUs detect and
utilize temporal spectrum holes over PU channels. Our results show
that compared to the existing approaches based on retransmission,
both PUs and SUs can achieve higher stable throughput, thanks to the
spectrum shaping effect of network coding.
\end{abstract}


\section{Introduction}
With the surge of a wide variety of wireless devices, the
traditional static spectrum allocation poses an obstacle for
efficient utilization of the limited spectrum resources. It is
difficult for new users to find communication opportunities whereas
the existing licensed or \textit{primary users} (PUs) barely utilize
the allocated spectrum to its full potential. According to the FCC
measurement, only $5\%$-$15\%$ on average has been used overall
\cite{Survey:FCC}. Such ``virtual scarcity'' in spectrum utilization
has spurred a wealth of interest in studying \textit{cognitive
radio} (CR) that provides \textit{secondary users} (SUs) with the
capability of dynamically adapting physical layer characteristics to
the available spectrum resources, thus enabling opportunistic access
to the primary spectrum \cite{thesis:CR}.

Cognitive radio goes beyond the conventional understanding of fixed
network resource allocation and enables the coexistence of multiple
user classes with varying priorities in a dynamic and hierarchical
spectrum sharing environment. Hence, it is important for the SUs to
capture temporal and spatial ``spectrum holes'' on PU channels,
thereby enabling opportunistic spectrum access. One grand challenge
in the design of CR networks is to discover spectrum holes for SUs
and distribute them efficiently in heavily-loaded systems. In
particular, a key functionality needed is the capability of sensing
the spectrum and opportunistically accessing it without causing
interference to the PUs. Spectrum holes appear in different ways,
over space, time, and frequency. Particularly, the traffic pattern
of the PUs directly determines the temporal spectrum availability
and possibly leaves room for SU transmissions whenever PUs do not
have any packet traffic to transmit.

In this paper, we consider a CR network with multiple PU channels,
each representing a PU subnetwork. Within each subnetwork, PU
packets are generated according to a stationary stochastic process.
A base station (BS) over the PU channel buffers randomly arrived
packets in its queue and multicasts them to a set of receiving PU
nodes over lossy wireless channels. As expected, PU channels are not
necessarily always busy and can be underutilized depending on packet
arrival and link rates. With this observation, we analyze how SUs
can discover these temporal spectrum holes (due to random packet
traffic and random channel conditions) by trading channel sensing
with data transmissions in a dynamic spectrum access environment.

Since SUs transmit on PU channels at idle instances of PUs, from a
holistic perspective it is important to increase the transmission
rates of PUs for given arrival rates, in order to enhance the
spectrum access opportunities of SUs. However, the random nature of
packet arrivals leads to stochastic and sporadic transmission
patterns over the PU channels and ``hides'' the temporal spectrum
holes from the SUs. We believe that it
is beneficial to introduce some tangible structure into PU transmissions in
the multicast setting with lossy communication links. To this end,
we leverage \textit{network coding} applied at PU channels to extend
spectrum holes for SUs, as well as to make them more predictable to
the SUs, i.e., network coding is used as a \textit{spectrum shaper}
on PU traffic.

Network coding is a novel networking paradigm that transforms the classical store-and-forward based
routing. By allowing intermediate nodes to code over the incoming
packet traffic, it is possible to improve the achievable throughput
to the min-cut capacity for general multicast communications
\cite{RYeung}. This coding diversity can be optimally realized by
linear network coding \cite{Li}, and distributed implementation is
possible through random linear network coding (RLNC)
\cite{TraceyHo_rnc} that improves the maximum flow rate to the
min-cut capacity with high probability. The throughput benefits of
network coding are not only possible in multi-hop operation but also
apply to single-hop broadcast channels \cite{Eryilmaz_Medard_IT08,
Medard_queueing, Tuan, Sagduyu_IT09} under both backlogged and
stochastic packet traffic.  Therefore, we can utilize network coding
to improve PUs' throughput rates, thereby extending temporal
spectrum resources for SUs.

Beyond this apparent throughput gain for PUs (and extension of
spectrum availability to SUs), we note that network coding further
introduces a predictive structure associated with PU transmissions.
Since PUs need to buffer batches of packets before coding them,
their transmissions become more predictable and therefore make it
easier for the SUs to discover the spectrum holes, whenever PUs
become idle. This use of network coding here is similar in spirit to
traffic shaping (e.g., leaky bucket \cite{ReneCruz}). Simply put,
traffic shapers buffer the incoming data and retransmit them over
time. The outgoing traffic resulted from the shaper is smoother and
more regular at the expense of additional delay involved through
buffering process. The same effect can be realized by network coding
while boosting the throughput of both PUs and SUs in CR networks
(without sacrificing delay performance in stable operation), and
this is the main focus of our study.

We formulate a spectrum shaping framework where each BS over a PU
channel first accumulates randomly arrived packets in its buffer and
then applies RLNC to combine them before multicasting the coded
packets to the receivers. Intuitively, when PUs use network coding,
the busy periods on each of the PU channels are lower-bounded by the
batch size of packets, and the idle periods are shaped based on the
process of accumulating the packets. Transitions between the idle
and busy periods become less frequent and more predictable due to
the buffering of packets and their batch-based transmissions.
Therefore, network coding applied by PUs reduces the need for
channel sensing and creates more space for SUs' data transmissions,
thereby improving SUs' throughput via the spectrum shaping effect,
compared to ARQ-based retransmission schemes.

We propose an adaptive channel sensing scheme where the SU sets a
timer for each PU channel that is sensed to be busy and does not
revisit it for a period of time. In this way, the SU tracks a
candidate list (we call this ``sensing list'' in subsequent
sections) of possibly idle PU channels and performs a two-stage
channel sensing: first, the SU senses the channels in the candidate
list (with each of them being chosen randomly and sensed until an
idle channel is detected) and then continues with the rest of the PU
channels, provided that the candidate list does not include any idle
channel. The timers for the PU channels form Markov chains that are
coupled through the sensing list size. We characterize the candidate
list evolution and compare the resulting throughput with random
channel sensing (which is carried out independently of the sensing
history). Significant throughput gains are attainable over ARQ when
the PUs apply network coding and the SU performs random channel
sensing. Further throughput gain is achieved when the SU performs
the two-stage adaptive channel sensing supported by network coding
over PU channels.

Our main contributions are three-fold:
\begin{enumerate}
\item We propose to leverage network coding as a spectrum shaper in a CR network, aiming to enhance the spectrum discovery for SUs.
\item We show that with network coding applied over PU channels only, both PU and SU's throughput are increased, compared to the case when ARQ-based retransmission schemes are used by the PUs.
\item We develop different sensing strategies for the SUs, and characterize the throughput attainable under random traffic, by balancing the tradeoffs between channel sensing and data transmissions for efficient spectrum access. The analysis is validated by simulation results.
\end{enumerate}

In related work on spectrum sensing, much work has been carried out
on exploiting the geographic and temporal properties of the signal
power (e.g., energy, cyclostationary signal characteristics,
interference and so on). Supportable communication regions, under
which the SUs can safely transmit without interfering with the PU
transmissions, were examined \cite{NeXt, SHaykin, LCao08, NChang08,
YChenIT08, YShi07, SHuang08, ASahai, Neely08}. In reactive spectrum
sensing, the SUs first sense the PU channels to detect the existence
of PUs, and then decide whether to transmit or not based on the
detection result \cite{swang10}. In proactive spectrum sensing, the
SUs predict future spectrum availability by exploring the history of
spectrum usage and switch to a different channel expected to be
available \cite{Heather_Proactive}. Our proposed method is a synergy
of these two approaches, in the sense that the SU senses PU channels
before transmission and prioritizes a candidate list of PU channels
that are expected to be idle, and starts sensing from this candidate list. At the MAC-layer, a partially
observable Markov decision process (POMDP) was used in \cite{myopic}
to model SUs' sensing process and to search for the optimal sensing order.
Also, a Markov decision process (MDP) framework was applied in,
e.g., \cite{mingyanliu, LifengLai}, to determine the optimal sensing
policy. Further, sensing period optimization was also studied (see
e.g.,\cite{KangShin}).

Most existing studies focused on SUs' performance, while little
special attention was paid to the holistic performance improvement
of both PUs and SUs. In contrast, our work here utilizes network coding
as a spectrum shaper to create a more predictable structure of
the spectrum holes, which simultaneously improves the throughput
performance of both PUs and SUs.

The rest of the paper is organized as follows. Section
\ref{sec:sysmodel} describes the system model, wherein a brief
summary of PU's stable throughput analysis is provided in Section
\ref{sec:PU}, for the two cases when PUs employ either ARQ or
network coding. In Section \ref{sec:SU}, we introduce the notion of
spectrum shaping via network coding and develop two (random and
adaptive) channel sensing schemes. Detailed analysis on random
channel sensing is provided in Section \ref{sec:NCrandom}, while
Section \ref{sec:NC2stage} extends the study to adaptive channel
sensing. In Section \ref{sec:numerical}, SU's throughput gains when
network coding is used by the PUs are evaluated. In particular, the
gain of adaptive channel sensing over random channel sensing is
demonstrated. Finally we conclude the paper in Section
\ref{sec:conclusion}.

\section{System Model \& Background} \label{sec:sysmodel}
\subsection{System Model}

We consider a CR network consisting of $N$ PU channels. We focus on
characterizing the spectrum shaping effect of network
coding, in terms of the structural properties of PU spectrum holes,
and developing efficient methods for SUs to discover and utilize the
available spectrum resources. For ease of exposition, in our study here
we consider one SU only (and the analysis readily extends to
multiple SUs, each independently discovering spectrum holes on PU
channels). Once PU spectrum holes are identified, the next step of
coordinating SU transmissions can be handled separately depending on
the choice of the medium access control mechanism, which we leave as a
future design objective.

In this study, we assume that each PU channel is associated with a
PU subnetwork consisting of one PU base station (BS) and $L$
receiving PU nodes (as shown in Fig. \ref{fig:sysmodel}). Within
each PU subnetwork, the BS multicasts data packets to the receiving
nodes. We assume that the time is slotted and synchronized among PU
channels. In each slot, packets arrive at any PU channel,
$j=1,...,N$, according to a stationary arrival process with a common
rate $\lambda$. PU channels are lossy, with erasure probability
$\varepsilon$. We assume independent erasures across channels and
time slots. For ease of exposition, we consider this homogeneous CR
network model, although the study can be carried over to general
packet arrival models and channel erasure rates.

\begin{figure}[tbh]
    \begin{center}
        \includegraphics [width = 0.4\textwidth] {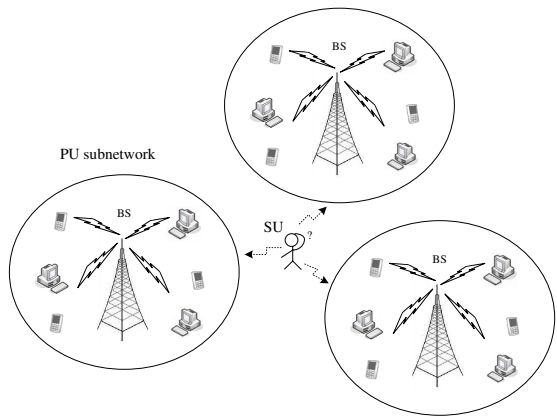}
        \caption{A cognitive radio network model.}
        \label{fig:sysmodel}
    \end{center}
\end{figure}

We assume that the transmission is carried
out using either network coding or ARQ. If network coding is used,
each BS accumulates a batch of $m$ packets, encodes them using
random linear network coding, and multicasts the coded packets to
the receivers. Once all receivers of a PU subnetwork decode the
entire batch of $m$ packets, the BS proceeds with the next batch of
$m$ packets, provided that there are enough buffered packets. If
not, the PU BS waits to accumulate the next batch of packets. Random
linear network coding is carried out in the Galois field $GF(q)$,
where $q$ is the finite field size. We assume a large value of $q$
so that with high probability (as specified in \cite{TraceyHo_rnc}),
each generated coded packet would be \textit{innovative} and each
receiver simply needs to receive exactly $m$ coded packets in order
to decode the whole batch of packets.

In contrast, if ARQ is used by the PUs, each BS multicasts individual packets one by one. The BS keeps retransmitting each packet (in uncoded plain form) until all receivers successfully receive it. Then, the BS proceeds with transmitting the next packet in the buffer.

\begin{figure}[tbh]
    \begin{center}
        \includegraphics [width = 0.4\textwidth] {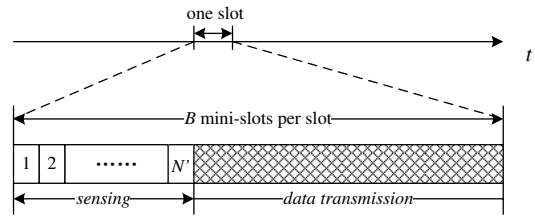}
        \caption{SU's slot structure for channel sensing and data transmission.}
        \label{fig:mini-slot}
    \end{center}
\end{figure}

As is standard, the SU opportunistically explores spectrum holes on
PU channels for data transmissions. We assume synchronization across
the SU and PUs. The packet transmission of a PU continues one slot
(or time frame) and each slot of the SU amounts to $B$
mini-slots, which are used either for sensing different PU channels
or transmitting packets. Clearly, at most $B$ channels can be sensed
per slot, i.e., the SU would be able to sense all PU channels when
needed only if $N\leq B$. Sensing is performed based on the
``sensing list'' $\mathcal{N}_t$ in every slot $t$, which consists
of the channels that are considered by the SU as ``possibly idle''
in this slot. When sensing, the SU picks a channel randomly and
uniformly from the list at a time without replacement. After channel
sensing, the SU transmits data packets during the rest of the time
slot, provided that an idle channel is detected; otherwise, the SU
waits till the next slot and repeats the procedure. Fig.
\ref{fig:mini-slot} illustrates the slot structure, where $N'$
points to the end of the sensing phase whenever the SU decides to
transmit.

It is clear that when PUs use network coding, the transmission is
performed on a batch basis with size $m$. On the other hand, when
ARQ is used by the PUs, the transmission is carried out per
individual packet. This major difference implies that network coding
not only extends the possible spectrum hole but also makes the
spectrum structure more ``regular,'' namely, easier for SUs to
predict when it is idle. With this insight, we should expect that
the SU keeps a shorter sensing list when PUs use network coding
compared to the case when ARQ is used. More specifically, when PUs
use network coding, we have
\begin{eqnarray}
N_t \leq N,~~~\forall~t,
\end{eqnarray}
and when PUs use ARQ,
\begin{eqnarray}
N_t = N,~~~\forall~t,
\end{eqnarray}
where $ N_t = |\mathcal{N}_t|$ is the size of the sensing list. In
particular, the sensing list size can be updated dynamically, when
network coding is used, and we will show in Section
\ref{sec:numerical} that this results in higher throughput beyond
random channel sensing. In the sequel, we will analyze different
strategies taken by the SU for channel sensing, for cases when PUs
use ARQ or network coding. Before that, we briefly demonstrate the
gain brought by network coding over the PU channels in the
following.

\subsection{PU's Throughput and Idle Probability} \label{sec:PU}
For broadcast erasure channels, it is well-known that network coding
reduces the completion time of data transmission compared to
retransmission schemes \cite{Eryilmaz_Medard_IT08},
\cite{Sagduyu_IT09}. For completeness, we characterize the stable
throughput and busy/idle periods of PUs for both cases when PUs use
network coding and ARQ in the following.

\subsubsection{When PUs use network coding}
Random network coding with large $q$ ensures that each of the $L$ PU
receivers can decode the original $m$ packets with high probability,
as long as it receives exactly $m$ coded packets
\cite{TraceyHo_rnc}. Let $T_{_{NC}}$ denote the completion time for
all receivers to successfully decode $m$ packets. Based on
\cite{Eryilmaz_Medard_IT08}, the expected time is given by
\begin{eqnarray} \label{eq:E_T_nc}
E[T_{_{NC}}]
   \hspace{-2.5mm}&=& \hspace{-2.5mm} m \hspace{-0.5mm} + \hspace{-0.5mm}\sum_{t=m}^{\infty} \left[ \hspace{-0.5mm}1 \hspace{-0.5mm}-\hspace{-0.5mm} \left(\hspace{-0mm}\sum_{a=m}^t \hspace{-0.5mm}{{a-1} \choose {m-1}} (1 \hspace{-0.5mm}-\hspace{-0.5mm}\varepsilon)^m \hspace{-0.5mm}\varepsilon^{a-m}\hspace{-0.5mm}\right)^L \hspace{-0.5mm}\right]. \nonumber \\
\end{eqnarray}

The stability condition of the PU queues is given by $\lambda < \eta_p^{NC}$, where
the maximum stable throughput $\eta_p^{NC}$ is
\begin{eqnarray}
\eta_p^{NC}  = \frac{1}{E[T_{_{NC}}]/m}.
\end{eqnarray}

From Little's theorem \cite{DataNetworks}, the idle probability of each PU BS can be obtained as
\begin{eqnarray}
P_{idle}^{NC} = 1 - \frac{\lambda E[T_{_{NC}}]}{m}.
\end{eqnarray}

\subsubsection{When PUs use ARQ}
For ARQ, the completion time of any individual packet is the number
of time slots necessary for successful reception of it at all
receivers. Let $T_{0,ARQ}$ denote this time. Note that ARQ is
equivalent to network coding with batch size $m=1$. From
(\ref{eq:E_T_nc}) with $m=1$, the expected service time is obtained
as
\begin{eqnarray} \label{eq:E_T_arq}
E[T_{0,ARQ}] =  \sum_{t=1}^ \infty \left(1 - \left(\sum_{a=1}^t (1 - \varepsilon)\varepsilon^{a-1}\right)^L \right) + 1.
\end{eqnarray}

The stability condition of the PU queues is given by $\lambda < \eta_p^{ARQ}$, where the maximum stable throughput $\eta_p^{ARQ}$ is
\begin{eqnarray}
\eta_p^{ARQ} = \frac{1}{E[T_{0,ARQ}]}.
\end{eqnarray}

It follows that the idle probability of the PU channels is
\begin{eqnarray}
P_{idle}^{ARQ} = 1 - \lambda E[T_{0,ARQ}].
\end{eqnarray}

As a result, we have $E[T_{_{NC}}] < mE[T_{0,ARQ}]$ and therefore
\begin{eqnarray}
\eta_p^{ARQ} &<& \eta_p^{NC}, \nonumber \\
P_{idle}^{ARQ} &<& P_{idle}^{NC}.
\end{eqnarray}

As expected, network coding increases the stable throughput for PUs
and provides the SU with extended availability of the primary
spectrum. Beyond this throughput gain, as we show in the
subsequent sections, network coding also shapes the spectrum and
increases the predictability of whether PU channels are idle or not,
which in turn reduces the channel sensing time by the SU, thus further improving the SU's
throughput.

\section{Spectrum Shaping via Network Coding} \label{sec:SU}
We start this section with a special case with a single PU channel,
and illustrate the basic idea of spectrum shaping. We then turn to
the general multi-channel scenario, which
is the main focus of the analysis in the sequel.

\subsection{The Case with a Single PU Channel}

When there is a single PU channel, the SU is able to track the
dynamics on the channel continuously by sensing it every slot.
Intuitively, such a strategy is plausible if the PU uses ARQ-based
transmission mechanism, since the packet arrivals are random and the
transmission is carried out on a per packet basis. Then, the SU
senses the channel using the first mini-slot in every slot, and
after sensing, if the channel is sensed to be busy, the SU backs off
and senses it again.

However, when network coding is used by the PU, the spectrum is better
``shaped'' in the sense that the transmission period $T_{_{NC}}$
becomes more regular and is bounded below by the batch size $m$.
Consequently, it is not necessary for the SU to sense the channel
every slot while not losing tractability on it. Instead, the SU
simply backs off for $m$ slots when the channel is sensed to be
busy, and starts sensing every slot after $m$ slots. Clearly, the SU
is able to recognize the starting time slot for \textit{every}
transmission period after it finds the channel idle in the first
place. Therefore, such a backoff scheme is accurate in predicting
the minimum completion time (i.e., $m$) while preventing the SU from
missing any potential spectrum holes, and this reduces the sensing
overhead without sacrificing the SU's throughput, compared to the case
when the PU uses ARQ.

\subsection{The Case with Multiple PU Channels}

When there are $N$ channels in the CR network, the SU is no longer
capable of keeping track of each channel all the time. The SU either
senses a subset of them before transmitting, or gives up channel
sensing for the current time slot (if none of the channels can be
idle). Intuitively speaking, the SU needs to find an idle channel as
soon as possible, while keeping the sensing cost low by exploiting
the underlying traffic structure. When ARQ is used by the PUs, the
PU channels experience fast idle-busy alternations on a per slot
basis. However, when network coding is used by the PUs, the
transitions between the idle and busy states would become slower on
the scale of $m$ slots. Therefore, if network coding is used by the
PUs, the SU would be better off by adaptively updating the sensing
list. Based on the relationship of the time spent for channel
sensing, namely sensing cost, and the achievable throughput, the SU
can take different strategies for updating the sensing list.

We develop the following two strategies for the SU:
\begin{itemize}
\item \textit{Random channel sensing under network coding and ARQ.}~~~
In every slot, the SU picks one channel randomly and uniformly from
the complete list (i.e., the list with all $N$ channels) and senses
it. If the channel is busy, the SU chooses another channel randomly
and uniformly in the next mini-slot and senses it. If the channel is
busy, the SU continues. Otherwise, the SU stops at the channel and
transmits on it using the rest of the slot. Loosely speaking, random
channel selection is more appropriate for ARQ, since the
alternations between idle and busy states are fast. Nevertheless, as
we show later, when PUs apply network coding, the SU's
throughput can be significantly improved, compared to the case when
ARQ is used.

\item \textit{Backoff-based adaptive sensing under network coding.}~~~

As pointed out before, when network coding is used by PUs,
the SU can predict ``slower'' transitions between PUs' idle and busy
states due to additional buffering and batch transmissions of coded
packets. Instead of keeping all $N$ channels in the sensing list,
the SU seeks for a shortened list while hoping to reduce the time it
takes to find an idle channel. In each time slot, the SU carries out
a two-stage sensing. First, the SU starts sensing PU channels
randomly picked from the sensing list $\mathcal{N}_t$ one by one,
and stops whenever an idle PU channel is detected. In the meantime,
SU backs off on channels sensed to be busy for $k$ slots and updates
the list $\mathcal{N}_t$ based on the channel sensing information.
The SU sets a timer for each PU channel it backs off and moves the
particular channel back to the candidate sensing list only after $k$
slots.

\hspace{4mm}If all channels in the first stage are found to be busy, the SU
proceeds to the second stage and randomly searches for an idle
channel in the backup list $\bar{\mathcal{N}}_t$, namely the list of
channels excluding $\mathcal{N}_t$, with size $|\bar{\mathcal{N}}_t|
= N-N_t$, until it finds one.\footnote{Note that the sensing
capability can be further improved by ordering channels in the
back-up list according to their timer values and giving priority to
those channels with smaller timer values in the sensing order.
However, this would increase the complexity significantly. Instead,
we consider random channel selection from the backup list in the
second stage.} In this stage, the SU moves the channel that it
senses idle back to the sensing list, but does not change the timers
of channels that are detected to be busy.

\hspace{4mm}The intuition behind this two-stage sensing approach is to give
higher sensing priority in the first stage to channels that are more
likely to be idle while seeking for an idle channel beyond the
priority list, if necessary. The backoff parameter $k$ can be
expressed as a function of the batch size $m$, namely $k = g(m), g:
\mathbb{Z}^+ \rightarrow \mathbb{Z}^+$. Without prior information
(i.e., when the SU visits a busy channel for the first time), on
average the SU would find the PU transmissions in the middle of the
batch service time. Therefore, we can intuitively let $k =
\frac{E[T_{_{NC}}]}{2}$, with $E[T_{_{NC}}]$ given in
(\ref{eq:E_T_nc}). For any channel, the SU should choose as the
backoff timer $k$ the average remaining time it believes for the PU
to complete the transmission of the current batch. We will evaluate
the effects of $k$ on the channel sensing and throughput performance
in Sections \ref{sec:NC2stage} and \ref{sec:numerical}.

\end{itemize}

\subsection{SU's Throughput}

Let $D_t$ represent the number of mini-slots used for sensing in
slot $t$ by the SU. Define a ``good slot'' to be one in which the SU
finds an idle channel and transmits. Correspondingly, a slot is
called ``bad'' if the SU fails in obtaining any transmission
opportunity therein. Let $\textbf{1}_t$ be the indicator random
variable indicating whether slot $t$ is good ($\textbf{1}_t=1$) or
bad ($\textbf{1}_t=0$). Since the SU can transmit data packets only
over $B-D_t$ mini-slots (whenever SU does not sense the channels)
and only if the channel sensing is successful (i.e., only if it
detects one idle PU channel), the SU's throughput can be obtained as
\begin{eqnarray}
\eta_s &=& \lim_{T_{tot} \rightarrow \infty} \frac{\sum_{t=1}^{T_{tot}} (B-D_t)\textbf{1}_t}{T_{tot}} \nonumber \\
       &=& E[(B-D_t)\textbf{1}_t] \nonumber  \label{eq:eta_s_2} \\
       &=& B p_{_r} - E[D_t \textbf{1}_t] \label{eq:eta_s},
\end{eqnarray}
where $T_{tot}$ is the time period under observation, $p_{_r} = \textrm{Pr}(\textbf{1}_t = 1)$, and (\ref{eq:eta_s_2}) follows from the ergodicity of the channel sensing process.

In the following, we focus on multiple PU channels and show how
network coding enables new opportunities for dynamic spectrum sensing
and improves the SU's throughput compared to ARQ, while reducing the sensing cost.

\section{Random Channel Sensing} \label{sec:NCrandom}

When the SU uses random channel sensing, it probes the channels one
by one randomly in each slot, until it finds an idle channel.

\subsection{When PUs use Network Coding}\label{subsec:maximum_effort}
Note that the SU cannot sense all $N$ channels when $B$ is smaller
than $N$ (even though it keeps the sensing list size unchanged as
$N_t = N$). Accordingly, we study two cases in the sequel: $B \geq
N$ and $B < N$.

When $B \geq N$, i.e., when the slot length can accommodate sensing
all channels, the probability that there exists at least one idle PU
channel is
\begin{eqnarray} \label{eq:pr}
p_{_r} &=& 1 - (1-P_{idle}^{NC})^N,
\end{eqnarray}
and, if an idle channel exists, the number of mini-slots used for
sensing is
\begin{eqnarray}
\textrm{Pr}(D_t = d | \textbf{1}_t = 1) &=& P_{idle}^{NC}(1-P_{idle}^{NC})^{d-1},
\end{eqnarray}
for $d \in \{1,...,N\}$. Based on (\ref{eq:eta_s}), the throughput of the SU can be derived as
\begin{eqnarray}
\eta_s^{NC}  \hspace{-2mm}&=&\hspace{-2mm} \Big(B-\frac{1-(1-P_{idle}^{NC})^N (1+N P_{idle}^{NC})}{P_{idle}^{NC}}\Big) \nonumber \\
             \hspace{-2mm}&& \times\big(1-(1-P_{idle}^{NC})^N\big).
\end{eqnarray}

On the other hand, when $B < N$, we shall consider only the first $B$ slots. Correspondingly, we have
\begin{eqnarray}
p_{_r} &=& 1-(1-P_{idle}^{NC})^B, \\
\textrm{Pr}(D_t = d | \textbf{1}_t =1) &=& P_{idle}^{NC}
(1-P_{idle}^{NC})^{d-1},
\end{eqnarray}
for $d \in \{1,...,B\}$. In this case, the SU's throughput is given
as
\begin{eqnarray}
\eta_s^{NC}  \hspace{-3mm}&=&\hspace{-3mm} \Big(B - \frac{1-(1-P_{idle}^{NC})^B(1+BP_{idle}^{NC})}{P_{idle}^{NC}}\Big) \nonumber \\
             \hspace{-3mm}&&\hspace{-3mm}\times
             (1-(1-P_{idle}^{NC})^B).
\end{eqnarray}

\subsection{When PUs use ARQ}
Similarly, we consider two regions as in the previous case. When $B \geq N$, we have
\begin{eqnarray*}
p_{_r} &=& 1 - (1-P_{idle}^{ARQ})^N, \\
\textrm{Pr}(D_t = d | \textbf{1}_t = 1) &=&
P_{idle}^{ARQ}(1-P_{idle}^{ARQ})^{d-1},
\end{eqnarray*}
for $d \in \{1,...,N\}$.

It follows that the SU's throughput is
\begin{eqnarray}
\eta_s^{ARQ} \hspace{-2mm}&=&\hspace{-2mm} \Big(B-\frac{1-(1-P_{idle}^{ARQ})^N (1+N P_{idle}^{ARQ})}{P_{idle}^{ARQ}}\Big) \nonumber \\ \hspace{-2mm}&& \times \big(1-(1-P_{idle}^{ARQ})^N\big).
\end{eqnarray}

On the other hand, when $B < N$, the throughput of the SU can be obtained as
\begin{eqnarray}
\eta_s^{ARQ} \hspace{-3mm}&=&\hspace{-3mm} \Big(B - \frac{1-(1-P_{idle}^{ARQ})^B(1+BP_{idle}^{ARQ})}{P_{idle}^{ARQ}} \Big)  \nonumber \\
             \hspace{-3mm}&&\hspace{-3mm} \times (1-(1-P_{idle}^{ARQ})^B).
\end{eqnarray}

\section{Backoff-based Adaptive Sensing} \label{sec:NC2stage}
\subsection{The Case with $B \geq N$} \label{subsec:NC2stage_Nsmall}

We start with the case $B \geq N$. In adaptive sensing, the SU
dynamically updates the size of the sensing list based on the
sensing results from the current slot. It follows that the sensing
list size, $N_t$, becomes a random variable in itself in this case.
For notational convenience, denote $p_n = \textrm{Pr}(N_t = n)$. To
quantify the SU's throughput, we consider two stages: sensing within
the sensing list $\mathcal{N}_t$ and beyond it within the backup
list $\bar{\mathcal{N}}_t$. Define $\textbf{1}_t^{(1)}$
(respectively $\textbf{1}_t^{(2)}$) as the indicator function
indicating whether there exists at least one idle channel in the
first (respectively second) stage. Clearly, only if the SU fails in
the first stage, i.e., all channels in the sensing list are found to
be busy, the SU will proceed into the second stage. Accordingly,
$p_{_r}$ can be obtained as the same in (\ref{eq:pr}), and the number
of mini-slots used for finding an idle channel, provided that there
exists at least one idle PU channel, is computed as
\begin{eqnarray} \label{eq:edt}
&&E[D_t\textbf{1}_t]= \sum_{n=0}^N\sum_{d=1}^n d ~\textrm{Pr} (D_t =d,\textbf{1}_t^{(1)}=1 | N_t =n) p_n \nonumber\\
\hspace{-4.5mm} &+&\hspace{-3.5mm} \sum_{n=0}^N \sum_{d=n+1}^N d~ \textrm{Pr}(D_t=d, \textbf{1}_t^{(1)}=0, \textbf{1}_t^{(2)}=1| N_t = n) p_n \nonumber\\
\hspace{-3.5mm}&=&\hspace{-3.5mm} \sum_{n=0}^N \frac{1-(1-P_{idle}^{NC})^n(1+nP_{idle}^{NC})}{P_{idle}^{NC}} \Big(1-(1-P_{idle}^{NC})^n \Big)p_n \nonumber\\
\hspace{-3.5mm}&+&\hspace{-3.5mm}  \sum_{n=0}^N \hspace{-0.5mm}\sum_{d=n+1}^N \hspace{-2mm}d P_{idle}^{NC}(1-P_{idle}^{NC})^{(d-n)-1}\Big(\hspace{-0.5mm}1-(1-P_{idle}^{NC})^{N-n}\hspace{-0.5mm}\Big)\nonumber\\
       \hspace{-3.5mm}&&\hspace{45mm}\times(1-P_{idle}^{NC})^n p_n.
\end{eqnarray}

The calculation of SU's throughput then follows from
(\ref{eq:eta_s}). However, it is observed that in order to
characterize the SU's throughput, we shall first characterize $p_n$,
the distribution of $N_t$. Note that once a PU channel is sensed to
be busy, the SU backs off on this channel by setting up a countdown
timer with an initial value $k$. The evolution of such a timer
follows a Markov chain, where states $0, ..., k$ correspond to the
countdown values of the timer, or equivalently, the remaining time
slots before the channel is moved back to the sensing list
$\mathcal{N}_t$. It is noted that the value of the timer decreases
from $k$ to $0$ at a rate of $1$ per slot. When it enters state $0$,
the corresponding channel is considered as a potential candidate in
the sensing list again, and it remains in the list as long as it is
sensed to be idle or it is simply not sensed. On the contrary, the
channel is removed from the sensing list in the next slot, provided
that it is sensed to be busy. In the meanwhile, if the SU enters the
second stage and senses the channels therein, the state of the idle
channel it finally finds changes to 0 in the next slot. Thereby, the
Markov chain for the timer should include transitions from states $i
= 2,...,k$ to state $0$, as depicted in Fig. \ref{fig:mc_k_ext}. Accordingly, the transition probabilities can be
expressed as
\begin{eqnarray} \label{eq:trans_matrix_extend}
p_{i,i-1} &=& 1 - p_b P_{idle}^{NC},~~i = 2,...,k, \nonumber \\
p_{i,0} &=& p_b P_{idle}^{NC},~~i=2,...,k, \nonumber \\
p_{1,0} &=& 1, \nonumber \\
p_{0,k} &=& p_s (1-P_{idle}^{NC}), \nonumber \\
p_{0,0} &=& p_s P_{idle}^{NC} + (1-p_s),
\end{eqnarray}
where $p_s$ is the probability that a PU channel in the sensing list is sensed in the first stage, and $p_b$ represents the probability that a PU channel in the backup list is sensed provided that all channels in the sensing list are found busy.

\begin{figure}[tbh]
    \begin{center}
        \includegraphics [width = 0.4\textwidth] {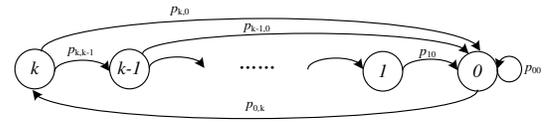}
        \caption{Markov chain for the timer of a given PU channel.}
        \label{fig:mc_k_ext}
    \end{center}
\end{figure}

Recall that the PU channels in both lists are randomly chosen and sensed, until an idle channel is detected. The Markov chains of timers
for different PU channels are coupled with each other through the size of the sensing list $\mathcal{N}_t$, which in turn depends on the states of the individual Markov chains, namely how many of them are in state $0$. An approximation is necessary to integrate the effects of all PU channels (similar approach has been taken in e.g., \cite{Bianch}, to analyze IEEE 802.11 backoff mechanism). In particular, we average the sensing probability of the first stage, $p_s$, and obtain that
\begin{eqnarray}
p_s = \sum_{n=1}^N \sum_{x=0}^{n-1} \prod_{x' = 0}^ {x-1}
\left(1-\frac{1}{n-x'}\right)(1-P_{idle}^{NC})^x \frac{1}{n-x}
p_{n}.
\end{eqnarray}

Along the same line, the sensing probability of the second stage, $p_b$, can be characterized as
\begin{eqnarray}
p_b \hspace{-1.5mm}&=&\hspace{-1.5mm} \sum_{n=0}^{N-1} (1-P_{idle}^{NC})^n p_{n} \sum_{y=0}^{l'-1} \prod_{y'=0}^{y-1} \left(1-\frac{1}{l'-y'}\right) \nonumber \\
\hspace{-1.5mm}&&\hspace{27mm}\times (1-P_{idle}^{NC})^y \frac{1}{l'-y},
\end{eqnarray}
where $l' = N-n$.

The stationary distribution for the states $i = 0,1,...,k$ in the Markov chain is given by
\begin{eqnarray} \label{eq:pi_extend}
\pi_0 &=& \pi_0 p_{0,0} + \pi_1 + \sum_{i=2}^k \pi_i p_{i,0}, \nonumber \\
\pi_{i-1} &=& \pi_{i} (1-p_b P_{idle}^{NC}),~~~i = 2,...,k,  \nonumber \\
\pi_k &=& \pi_0 p_{0,k}, \nonumber \\
\sum_{i=0}^k \pi_i &=& 1.
\end{eqnarray}

Based on (\ref{eq:trans_matrix_extend})-(\ref{eq:pi_extend}), $\pi_0$, the probability that any given PU channel is in the sensing list $\mathcal{N}_t$, can be computed as
\begin{eqnarray} \label{eq: pi0_ext}
\pi_0 = \Bigg(1 + p_s(1-P_{idle}^{NC}) \nonumber \hspace{45mm}\\
        \times \Big(1+\frac{(1-p_b P_{idle}^{NC})(1-(1-p_b P_{idle}^{NC})^{k-1})}{p_b P_{idle}^{NC}}\Big)\Bigg)^{-1}.
\end{eqnarray}

It follows that the size of the sensing list, $N_t$, has a binomial distribution with parameter $\pi_0$, i.e., $p_n = \textrm{Pr}(N_t = n)$ is expressed as
\begin{eqnarray} \label{eq:fixedpoint_pn}
p_n = {N \choose n} (\pi_0)^n (1-\pi_0)^{N-n}.
\end{eqnarray}

Let $\textbf{p} = [p_0, p_1,...,p_N]$. We note that (\ref{eq:fixedpoint_pn}) consists of a fixed point equation of form $\textbf{p} = T(\textbf{p})$ for $\textbf{p}$. Since $T: [0,1]^{N+1} \rightarrow [0,1]^{N+1}$ is a continuous mapping on a compact space, there exists a solution to $\textbf{p} = T(\textbf{p})$. Furthermore, $\textbf{p}$ consists of the stationary distributions of the positive recurrent Markov chain that represents the overall $N$ PU channel states (it has a finite number of states and it is commutative), it follows that there exists a unique solution to (\ref{eq:fixedpoint_pn}).

\subsection{The Case with $B < N$}
When $B < N$, the modeling of Markov chains and corresponding
analysis follow the same line as in the case with $B \geq N$. In particular, we
have that
\begin{eqnarray} \label{eq:pr_2}
p_{_r} \hspace{-3.5mm}&=&\hspace{-3.5mm} \sum_{n=0}^B \textrm{Pr}(\textbf{1}_t = 1 |N_t = n)p_n + \hspace{-1.5mm}\sum_{n=B+1}^N \textrm{Pr}(\textbf{1}_t = 1 |N_t = n)p_n\nonumber\\
       \hspace{-3.5mm}&=&\hspace{-3.5mm} \hspace{-0.5mm}\sum_{n=0}^B \hspace{-1mm}\Bigg(\hspace{-0.5mm} \Big(\hspace{-0.5mm} 1 \hspace{-0.5mm}-\hspace{-0.5mm}(\hspace{-0.5mm}1\hspace{-0.5mm}-\hspace{-0.5mm}P_{idle}^{NC}\hspace{-0.5mm})^n\hspace{-0.5mm}\Big) \hspace{-0.5mm}+\hspace{-0.5mm} ( \hspace{-0.5mm}1\hspace{-0.5mm}-\hspace{-0.5mm}P_{idle}^{NC}\hspace{-0.5mm})^n \hspace{-0.5mm}\Big(\hspace{-0.5mm}1-(1-P_{idle}^{NC})^{B-n}\hspace{-0.5mm}\Big)\hspace{-0.5mm}\Bigg) \nonumber\\
       \hspace{-3.5mm}&&\times p_n + \hspace{-1mm} \sum_{n=B+1}^N \Big(1-(1-P_{idle}^{NC})^B\Big)p_n,
\end{eqnarray}
and
\begin{eqnarray} \label{eq:edt2}
\hspace{-3.5mm}&&\hspace{-3.5mm}E[D_t\textbf{1}_t] \nonumber \\
\hspace{-3.5mm}&=&\hspace{-3.5mm} \sum_{n=0}^B E[D_t \textbf{1}_t |N_t =n]p_n  + \hspace{-1.5mm}\sum_{n=B+1}^N \hspace{-2mm} E[D_t \textbf{1}_t |N_t =n]p_n \nonumber\\
\hspace{-3.5mm}&=&\hspace{-3.5mm} \sum_{n=0}^B \Bigg(\sum_{d=1}^n d P_{idle}^{NC} (1-P_{idle}^{NC})^{d-1} \Big(1-(1-P_{idle}^{NC})^n\Big) \nonumber \\
\hspace{-4.5mm}&& + \sum_{d=n+1}^B d P_{idle}^{NC}(1-P_{idle}^{NC})^{(d-n)-1}(1-P_{idle}^{NC})^n \nonumber \\
\hspace{-4.5mm}&& \hspace{30mm} \times (1-(1-P_{idle}^{NC})^{B-n}) \Bigg) p_n \nonumber\\
\hspace{-3.5mm}&+&\hspace{-3.5mm} \sum_{n=B+1}^N \sum_{d=1}^B dP_{idle}^{NC}(1-P_{idle}^{NC})^{d-1} \Big(1-(1-P_{idle}^{NC})^B\Big) p_n,\nonumber\\
\end{eqnarray}
where $p_n$ is given by (\ref{eq:fixedpoint_pn}), with $p_s$ and $p_b$ changed to
\begin{eqnarray}
p_s \hspace{-1.5mm}&=&\hspace{-1.5mm} \sum_{n=1}^N \sum_{x=0}^{x_{_0}} \prod_{x' = 0}^ {x-1} \left(1-\frac{1}{n-x'}\right)(1-P_{idle}^{NC})^x \frac{1}{n-x} p_{n}, \nonumber \\
p_b \hspace{-1.5mm}&=&\hspace{-1.5mm} \sum_{n=0}^{B-1} (1-P_{idle}^{NC})^n p_{n} \sum_{y=0}^{(\hspace{-0.7mm}B-n)-1} \prod_{y'=0}^{y-1} \left(\hspace{-0mm}1-\frac{1}{l'-y'}\hspace{-0mm}\right)\nonumber \\
\hspace{-1.5mm}&&\hspace{25mm}\times(1-P_{idle}^{NC})^y \frac{1}{l'-y},
\end{eqnarray}
for $x_{_0} = \min(B,n)-1$ and $l' = N-n$. The throughput of SU can
be characterized based on (\ref{eq:eta_s}) for $p_{_r}$ given by
(\ref{eq:pr_2}) and $E[D_t\textbf{1}_t]$ given by (\ref{eq:edt2}).

\subsection{Prediction Accuracy of Spectrum Opportunities} \label{subsec:accuracy}

In the backoff-based sensing strategy, the SU predicts the spectrum holes by adaptively updating the sensing list. As can be noted from the analysis above, the calculation of SU's throughput heavily hinges on the distribution of the sensing list size, $N_t$, where the key parameter is the idle probability of the PU channels. Clearly, in the actual underlying system, this probability is given by $P_{idle}^{NC}$, while in the \textit{predicted} system built from the adaptive sensing strategy, the expected idle probability on the PU channels is $\pi_0$ given by (\ref{eq: pi0_ext}). In order to examine the prediction accuracy, we define the following $\mathcal{L}_1$ distance $\delta$ to quantify the difference between the two:
\begin{eqnarray}
\delta = |\pi_0 - P_{idle}^{NC}|.
\end{eqnarray}

Intuitively, the smaller the distance $\delta$, the more accurate
the prediction is. To get a more concrete sense, we plot in Fig. \ref{fig:Pidle_pi0_e} and
\ref{fig:abs_diff_k} some examples on the comparison of $\pi_0$ and
$P_{idle}^{NC}$, where we set $L=20, m = 8, \lambda = 0.4,
\varepsilon = 0.2$ and $k = 4$ as the default values. As can be seen
from Fig. \ref{fig:Pidle_pi0_e}, the prediction $\pi_0$ closely
tracks the idle probability $P_{idle}^{NC}$ of the actual system for
different erasure probabilities $\varepsilon$, indicating the
robustness of channel tracking against channel variations. On the
other side, as Fig. \ref{fig:abs_diff_k} demonstrates, when the backoff parameter $k$ increases, the difference
first sharply shrinks and then increases slowly after $k$ reaches a
certain value. This points to an optimal backoff parameter:
\begin{eqnarray}
k^* = \min_{k} |\pi_0 - P_{idle}^{NC}| \label{eq:optimal_k},
\end{eqnarray}
for capturing the spectrum holes. Intuitively, if $k$ is
chosen to be smaller than the optimal one, the sensing list would be
longer than necessary with redundant PU channels that are actually
busy. On the other hand, if $k$ is greater than
(\ref{eq:optimal_k}), the SU tends to perform a conservative sensing
policy with a shorter list of candidate channels to be sensed. 

\begin{figure}[tbh]
    \begin{center}
        \includegraphics [width = 0.45\textwidth] {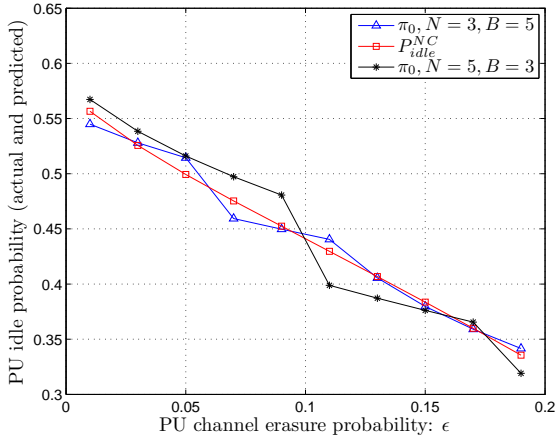}
        \caption{A comparison of the actual and predicted idle probability of PU channels in adaptive sensing with different PU erasure probabilities.}
        \label{fig:Pidle_pi0_e}
    \end{center}
\end{figure}

\begin{figure}[tbh]
    \begin{center}
        \includegraphics [width = 0.45\textwidth] {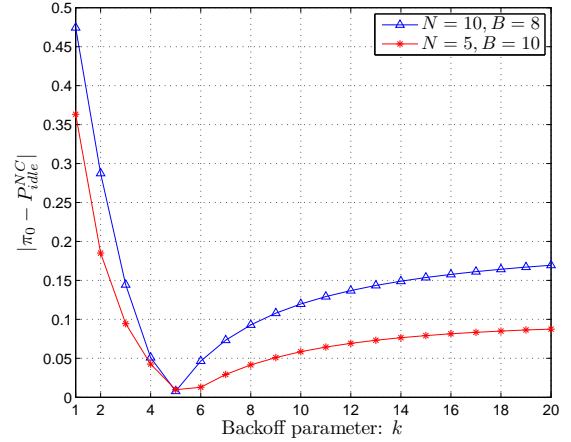}
        \caption{Difference between the actual and predicted idle probability of PU channels in adaptive sensing with different backoff parameters.}
        \label{fig:abs_diff_k}
    \end{center}
\end{figure}

\section{Performance Evaluation} \label{sec:numerical}
In this section, we illustrate, via numerical examples, the
throughput gain of the SU when PUs use network coding. Fig. \ref{fig:arq_ncR_e}
and \ref{fig:arq_ncR_lambda} show that compared to ARQ, the SU's
throughput is greatly increased when PUs use network coding and the
SU employs random channel sensing. Clearly, with
variations in the parameters, $\varepsilon$ and $\lambda$, the SU's
throughput improves with network coding, for both cases with $B \geq
N$ and $B < N$. Besides, as the values of these parameters increase,
the gain also increases.

Next, we show that by employing the adaptive sensing strategy, the
SU's throughput can be further improved by almost $15\%$, as Figs.
\ref{fig:percent_lambda}-\ref{fig:percent_k} indicate, where the
default values of parameters are $L = 20, m = 2, \lambda = 0.4,
\varepsilon = 0.2$ and $k = 2$. The two-stage adaptive sensing
strategy provides further gain over the random sensing mechanism,
over the entire region of parameter variations (in $m$, $\lambda$,
$\varepsilon$, and $k$). Fig. \ref{fig:percent_lambda} and
\ref{fig:percent_e} show that the gain increases with the PU arrival
rate and channel erasure probability, and as illustrated in Fig.
\ref{fig:percent_m}, the gain strongly depends on the network coding
batch size as well. Moreover, Fig. \ref{fig:percent_k} shows that
the backoff parameter can be further adjusted by the SU to improve
the gain of adaptive sensing.

We have a few more remarks on the above results.
As either $\lambda$ or $\varepsilon$ increases, the idle
probability of the PU channels decreases and the adaptive sensing
scheme, in which the SU makes more use of system information,
reveals more gain over the random sensing scheme, where
$P_{idle}^{NC}$ dominates the performance. Differently, if $m$
increases, the service rate of the PUs increases to the min-cut
capacity given by $1-\varepsilon$ \cite{Eryilmaz_Medard_IT08}. This
increases the idle probability of the PU channels for fixed arrival
rates and shrinks the operational difference between the random and
adaptive sensing strategies. Finally, we observe that optimizing
over the backoff parameter $k$ improves the performance of the
adaptive sensing strategy, as has also been observed in Section
\ref{subsec:accuracy}. On one hand, we have $\pi_0 = 1$ for $k=0$,
i.e., the random sensing strategy serves as a special case of the
adaptive sensing scheme for $k=0$. On the other hand, as $k$ increases, 
almost all PU channels are included in the backup list and
the adaptive sensing strategy performs again closely to the random
sensing scheme. Therefore, we expect an optimal $k$ with an
intermediate value to yield the highest performance gain.

\begin{figure}[tbh]
    \begin{center}
        \includegraphics [width = 0.49\textwidth] {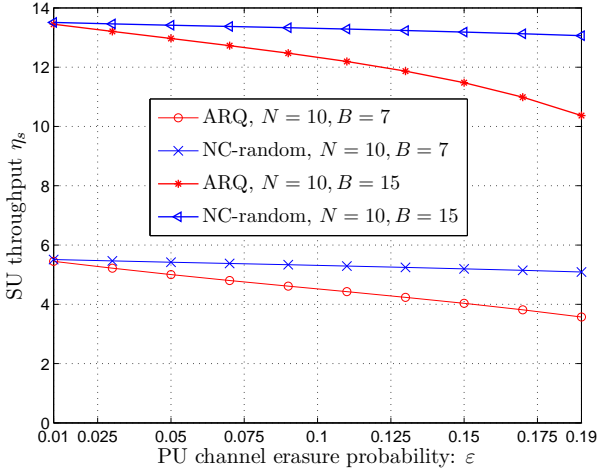}
        \caption{Throughput comparison: ARQ vs NC with random sensing for different PU erasure probabilities.}
        \label{fig:arq_ncR_e}
    \end{center}
\end{figure}
\begin{figure}[tbh]
    \begin{center}
        \includegraphics [width = 0.49\textwidth] {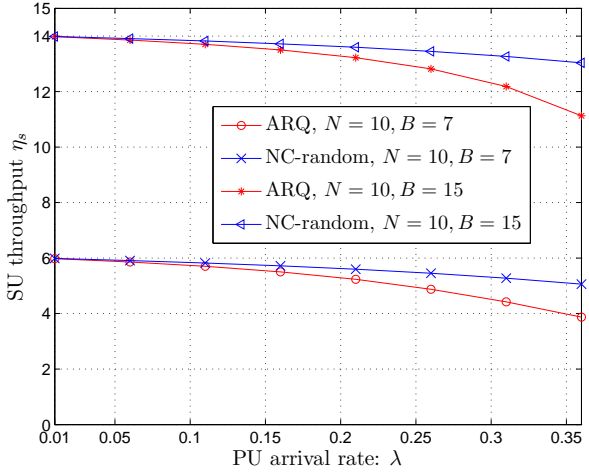}
        \caption{Throughput comparison: ARQ vs NC with random sensing for different PU arrival rates.}
        \label{fig:arq_ncR_lambda}
    \end{center}
\end{figure}

\begin{figure}[tbh]
    \begin{center}
        \includegraphics [width = 0.5\textwidth] {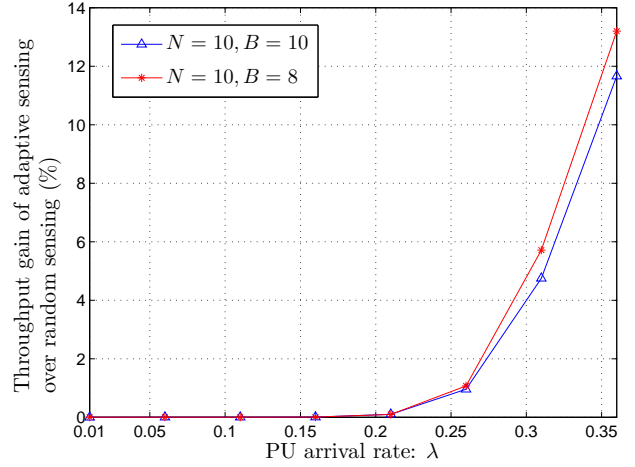}
        \caption{Gain of adaptive sensing over random sensing for different PU arrival rates.}
        \label{fig:percent_lambda}
    \end{center}
\end{figure}
\begin{figure}[tbh]
    \begin{center}
        \includegraphics [width = 0.5\textwidth] {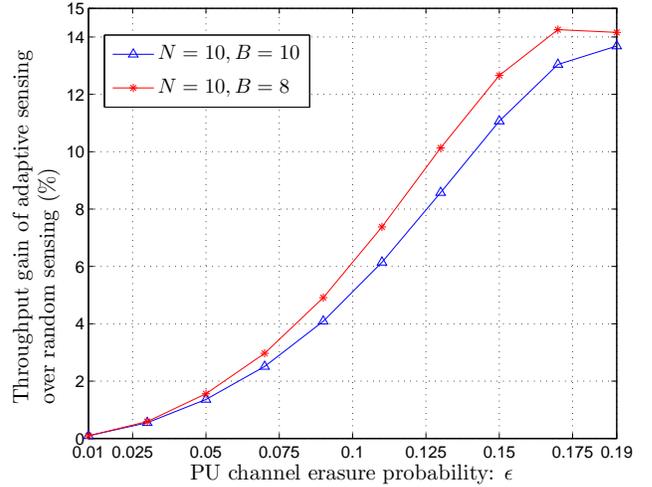}
        \caption{Gain of adaptive sensing over random sensing for different PU erasure probabilities.}
        \label{fig:percent_e}
    \end{center}
\end{figure}
\begin{figure}[tbh]
    \begin{center}
        \includegraphics [width = 0.48\textwidth] {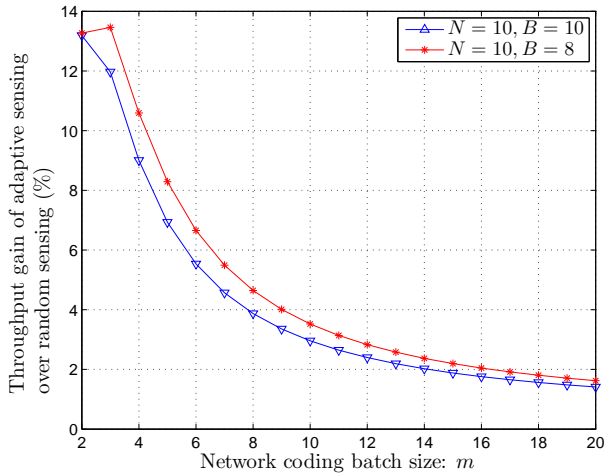}
        \caption{Gain of adaptive sensing over random sensing for different network coding batch sizes.}
        \label{fig:percent_m}
    \end{center}
\end{figure}
\begin{figure}[tbh]
    \begin{center}
        \includegraphics [width = 0.47\textwidth] {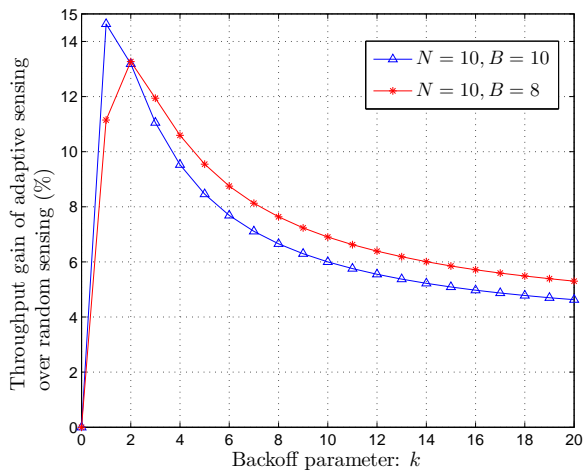}
        \caption{Gain of adaptive sensing over random sensing for different backoff parameters.}
        \label{fig:percent_k}
    \end{center}
\end{figure}

Finally, we take a closer look at the
adaptive sensing scheme. As pointed out in Section
\ref{subsec:NC2stage_Nsmall}, we approximate the sensing
probabilities $p_s$ and $p_b$ in the Markov chain analysis and compute
the SU's throughput accordingly. To validate this approximation, we
perform Monte Carlo simulations over $10^5$ trials and compare them
with the numerical evaluation of the analysis in Figs.
\ref{fig:sim_num_lambda}-\ref{fig:sim_num_k}. The default parameters
are $L=20, m =5, \lambda = 0.4, \varepsilon = 0.1$ and $k =2$. As
can be seen from the figures, the numerical results from the
analysis match with simulations under variations in all parameters.
On average, the maximum difference between the numerical and
simulation results is less than $3\%$, indicating that the
approximation in the analysis of adaptive sensing performs closely
to the real system implementation.

\begin{figure}[tbh]
    \begin{center}
        \includegraphics [width = 0.48\textwidth] {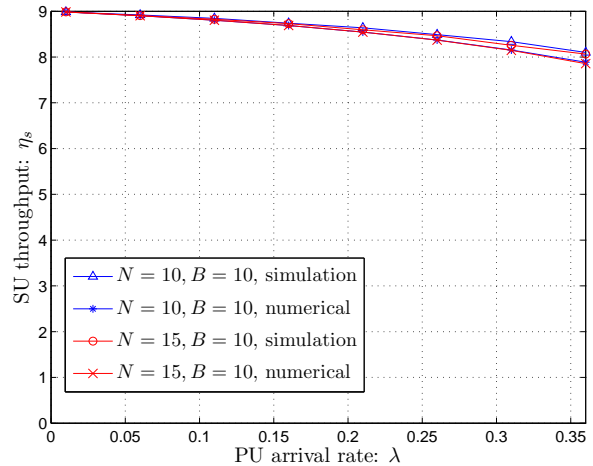}
        \caption{Comparison of simulation and numerical results for different PU arrival rates, when adaptive sensing is used.}
        \label{fig:sim_num_lambda}
    \end{center}
\end{figure}
\begin{figure}[tbh]
    \begin{center}
        \includegraphics [width = 0.48\textwidth] {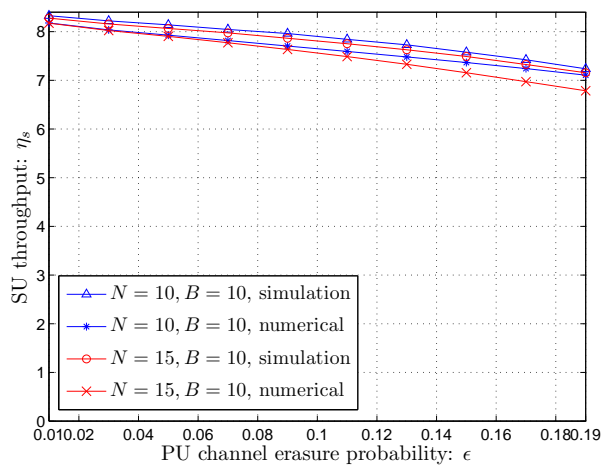}
        \caption{Comparison of simulation and numerical results for different PU erasure probabilities, when adaptive sensing is used.}
        \label{fig:sim_num_e}
    \end{center}
\end{figure}
\begin{figure}[tbh]
    \begin{center}
        \includegraphics [width = 0.5\textwidth] {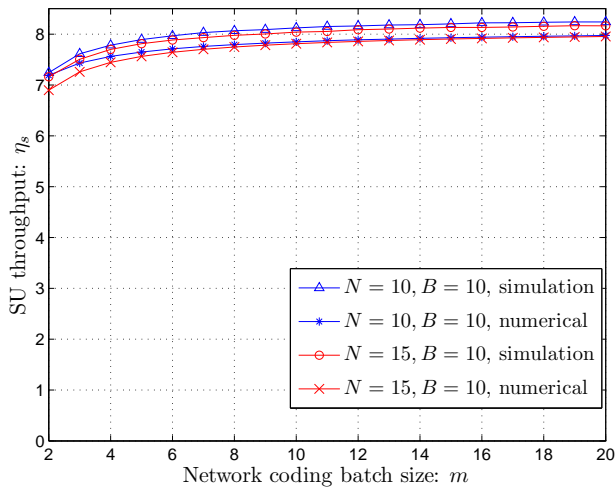}
        \caption{Comparison of simulation and numerical results for different network coding batch sizes, when adaptive sensing is used.}
        \label{fig:sim_num_m}
    \end{center}
\end{figure}
\begin{figure}[tbh]
    \begin{center}
        \includegraphics [width = 0.5\textwidth] {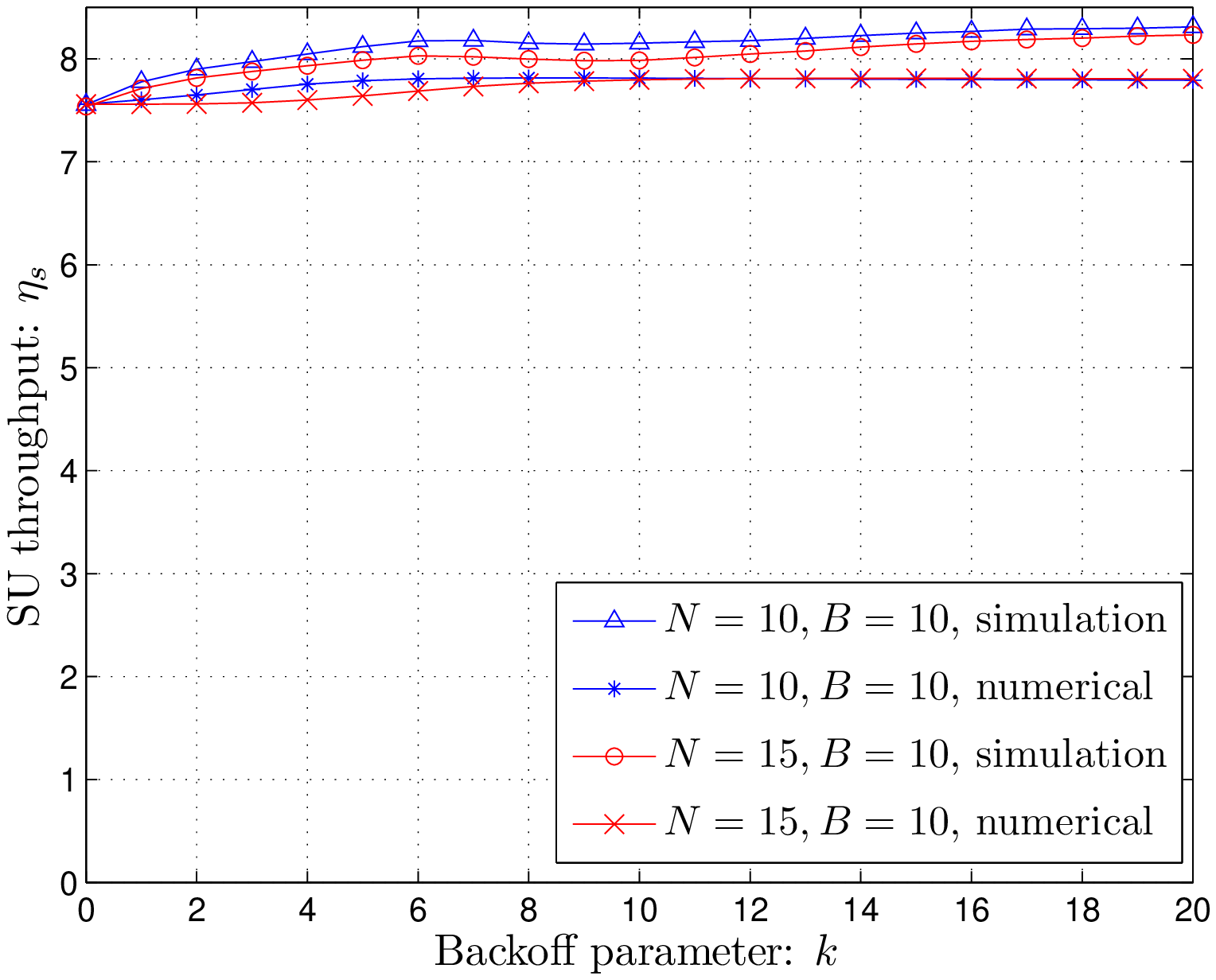}
        \caption{Comparison of simulation and numerical results for different backoff parameters, when adaptive sensing is used.}
        \label{fig:sim_num_k}
    \end{center}
\end{figure}

\section{Conclusion} \label{sec:conclusion}

We considered a CR network with $N$ PU channels and one SU, where
each PU transmits packets to multiple receivers over lossy wireless
channels via ARQ or network coding. Viewing network coding as a
spectrum shaper, we showed that it increases the spectrum
availability for the SU and offers a more predictive structure to
the PU spectrum, i.e. it improves the SU's prediction of spectrum
holes on PU channels. Based on the spectrum shaping effect of
network coding, we developed different sensing strategies for the
SU, where adaptive channel sensing is carried out by dynamically
updating the list of the PU channels that are predicted by the SU to
be idle. Our analysis and numerical results showed that compared to
retransmission, both PU and SU's throughput can be improved when PUs
apply network coding instead of ARQ, and the SU can further improve
this gain by applying adaptive channel sensing (based on sensing
history to reflect the PU traffic).

Future work is needed to quantify the gain of dynamic spectrum access for other channel models and to integrate medium access control for multiple SUs applying dynamic channel sensing with network coding. In particular, it is of great interest to study multi-hop CR networks where network coding has potential to offer more opportunities for PUs to improve their throughput. We expect that our initial steps here open a new avenue for SUs to discover spatial and temporal spectrum holes via spectrum shaping effects of network coding.

\nocite{*}
\bibliographystyle{IEEEtran}

\bibliography{cr_nc_arcv_ref}

\end{document}